\documentclass[referee]{aa} 
\usepackage{graphicx}
\begin{document}
   \title{High contrast experiment of an AO-free coronagraph 
with a checkerboard pupil mask}


   \author{K. Enya\inst{1},
           L. Abe\inst{2},
           S. Tanaka\inst{1,3},
           T. Nakagawa\inst{1},
           K. Haze\inst{1,4},
           T. Sato\inst{5},
           T. Wakayama\inst{5}
          }

\institute{Department of Infrared Astrophysics, Institute of Space and Astronautical Science, Japan Aerospace Exploration Agency, Yoshinodai 3-1-1, Sagamihara, Kanagawa 229-8510, Japan
\and
Optical and Infrared Astronomy Division \& Extra-solar Planet Project Office, National Astronomical Observatory, Osawa 2-21-2, Mitaka, Tokyo 181-8588, Japan 
\and
Department of Physics, Graduate School of Science, University of Tokyo, Hongo 7-3-1, Bunkyo-ku, Tokyo 113-0033, Japan 
\and
Department of Space and Astronautical Science, The Graduate University for Advanced Study, Yoshinodai 3-1-1, Sagamihara, Kanagawa 229-8510, Japan 
\and
Nanotechnology Research Institute, Advanced Industrial Science and Technology, Umezono 1-1-1, Tsukuba, Ibaraki 305-8568, Japan
}


 \date{Received ; accepted }

 
  \abstract
   {A high contrast coronagraph is expected to provide one of the promising ways to directly observe extra-solar planets. We present the newest results of our laboratory experiment investigating ``rigid'' coronagraph with a binary shaped checkerboard pupil mask, which should offer a highly stable solution for telescopes without adaptive optics (AO) for wavefront correction in space missions.}
   {The primary aim of this work was to study the stability of the coronagraph, and to demonstrate its performance without adaptive wavefront correction. Estimation of both the raw contrast and the gain of the point spread function (PSF) subtraction were needed. The limiting factor of the contrast was also important.
}
   {A binary shaped pupil mask of a checkerboard type has been designed. The mask, consisting of an aluminum film on a glass substrate, was manufactured using nano-fabrication techniques with electron beam lithography. Careful evaluation of coronagraphic performance, including PSF subtraction, was carried out in air using the developed mask.
}
   {A contrast of $6.7 \times 10^{-8}$ was achieved for the raw coronagraphic image by areal averaging of all of the observed dark regions. Following PSF subtraction, the contrast reached  $6.8 \times 10^{-9}$. Speckles were a major limiting factor throughout the dark regions of both the raw image and the PSF subtracted image. 
}
   {A rigid coronagraph with PSF subtraction without AO is a useful method to achieve high contrast observations. Applications of a rigid coronagraph to a Space Infrared telescope for Cosmology and Astrophysics (SPICA) and other platforms are discussed.
}

   \maketitle
%

\section{Introduction}

Direct detection and spectroscopy of extra-solar planets is essential to understand how planetary systems were born, how they evolve, and, ultimately, to find biological signatures on these planets. The enormous contrast in luminosity between the central star and a planet presents the primary difficulty in the direct observation of extra-solar planets. The typical contrast is $\sim$10-10at visible light wavelengths and $\sim$10-6 in the mid-infrared wavelength region (Burrows et al. 2004). A coronagraph, at first evaluated for the solar observation (Lyot 1936), could provide special optics to reduce the contrast.

Space telescopes have an advantage as platforms for high contrast coronagraphs because they are free from air turbulence. In space telescopes, the wavefront error (WFE) caused by imperfections in the optics is an important limiting factor in the contrast of coronagraph. Subtraction of the point spread function (PSF) is useful in canceling stable WFEs and achieving a higher contrast than the raw contrast of a coronagraph. Therefore, one of the promising ways to obtaining a high contrast observation is the combination of PSF subtraction with a ``rigid'' coronagraph, which does not include an adaptive wavefront correction system involving many actuators and is developed to be sufficiently precise and stable. Recent concepts such as the Super-Earth Explorer Coronagraphic Off Axis Space Telescope (SEE-COAST) proposal to the European Space Agency (ESA Cosmic Vision 2015-2020) has a baseline configuration which does not make any use of adaptive correction (Schneider et al. 2006). It relies on high precision optics polishing, as well as on the global stability of the telescope, for a goal detection contrast of 10-8 in the visible. The advantages of such a rigid coronagraph are simplicity, compactness, light weight, high stability and reliability, whilst adaptive wavefront correction would be another very useful method for a high contrast coronagraph (Trauger \& Traub 2007; Belikov et al. 2006).

In this paper, we present the results of a laboratory experiment to devise a rigid coronagraph in air, without adaptive wavefront correction, together with a discussion of the application of such a rigid coronagraph for the Space Infrared telescope for Cosmology and Astrophysics (SPICA) and other platforms.

\section{Experiment}

Enya et al. (2007a, hereafter Paper I) presented the results from the first laboratory experiment with a coronagraph using a binary shaped pupil mask of the checkerboard type, with visible light. Paper I refers to the manufacture of two masks (Mask 1 and Mask 2) on glass substrates. The contrast required in the design of Mask 1 and Mask 2 was 1.0 $\times$10-7. The achieved average contrast values, measured using the Mask 1 and Mask 2, were $2.7 \times 10^{-7}$ and $1.1 \times 10^{-7}$, respectively.

Considering the satisfactory results in Paper I, we again adopted a coronagraph employing a shaped pupil mask of the checkerboard type for this work, because it is simple, essentially achromatic, and relatively insensitive to pointing errors with the telescope (Jacquinot \& Roizen-Dossier 1969; Spergel 2001; Green et al. 2004; Vanderbei et al. 2004; Tanaka et al. 2006; Kasdin et al. 2003).

We developed a new checkerboard mask (Mask 3), in which the required contrast in the design was 1.0 $\times$10-10, representing the value required for the direct observation of extra-solar terrestrial planets. Optimization of the mask shape was performed using the LOQO solver presented by Vanderbei (1999). Figure 1 shows the solution of design in which required contrast, inner working angle (IWA), and outer working angle (OWA) were $1.0 \times 10^{-10}$, 3.7 $\lambda/D$, and 35 $\lambda/D$, respectively. Four dark regions (DRs) were named DR 1, DR 2, DR 3, and DR 4, as shown in Fig. 1b.

Fabrication of Mask 3 was, in principle, the same as shown in Paper I. Mask 3 consisted of an aluminum film with 100 nm thickness on a BK7 substrate and was manufactured using nano-fabrication technology at the National Institute of Advanced Industrial Science and Technology (AIST) in Japan. An anti-reflection coating was applied to both sides of the substrate. Figure 2a shows Mask 3, constructed on the substrate. The diameter of circumscribed circle around the transmissive region was 2 mm. Figure 2b is a photograph of the fabricated Mask 3 by a microscope observing transmitted light through the mask. Four small defects (holes) were found in the alminium film coating.

The configuration of the experiment is shown in Fig. 3. All the experimental optics were located in a darkened clean-room. Cleaned air flow was provided during the setting-up and measurements. The optics configuration is similar to that described in Paper I, except in a few points; we adopted a defect-covering mask to obscure light transmission through the defects shown in Fig. 2b. The defect-covering mask consisted of a 20 nm thick stainless sheet. The square holes in the covering mask were made by electoral discharge machining. The size of the square hole of the covering mask corresponds to the red square shown in Fig. 2b. The covering mask was attached to a rigid holder with an optical baffle. The unit comprising the covering mask, the holder and the baffle was mounted on a 5 axis adjustable stage. The entire unit was colored black. The unit was carefully positioned on the center of the checkerboard mask by observing the pupil image using transmitted light. The unit is set bring the covering mask into contact with the glass substrate of the checkerboard mask using a spacer of one layer of polyimide film tape. As the result, the four defects, shown in Fig. 2b, were covered, although the flexibility of optical configuration was vastly reduced because it was not easy to move the covering mask unit and the checkerboard mask whilst maintaining precise alignment. The checkerboard mask of aluminium was highly reflective, so a kick-off mirror was used to discard the light reflected by the checkerboard mask. A black box with inner baffles was used to catch and retain the reflected light.

For the imaging of the core of a coronagraphic PSF, a combination of several exposure times (0.03, 0.1, 1.0, 10 s) and the presence or absence of a neutral density filter, with an optical density of 2, were employed to achieve high dynamic range measurements. After each imaging, the laser source was turned off and a ``dark frame'' measurement was taken for each configuration, with the same exposure time and same optical density filter. The dark frame was subtracted from the image with the laser light on. The scaling by the exposure time and optical density allowed smooth profile of the core to be obtained (as in Fig. 6). The dark region of the coronagraphic image was observed with an 1800 s exposure. For the imaging of the dark region, a square-hole mask was set in front of the window of the CCD camera to prevent the flux that was not from the dark region entering the camera. The dark frame was taken with an 1800 s exposure after the each imaging of the dark region and then subtracted from the image with the laser on. The CCD was cooled and stabilized at 0 $^\circ$C throughout the experiment.

\section{Results and discussion }

The observed raw coronagraphic images and the result of the PSF subtraction are shown in Figs. 4 and 5, respectively. Figure 6 shows the profiles of the coronagraphic images obtained by measurement, together with an Airy profile and designed profile. The image and profiles of the core of the coronagraphic PSF are consistent with those expected from theory. In Fig. 4a, the horizontally spreading pattern between DR 1 and DR 2 is brighter than the pattern between DR 2 and DR 3. This asymmetry is caused by the CCD readout and strong saturation at the PSF core. 

The relative intensity in most of the area of the dark region was less than 10-6, as shown in Fig. 4b. The areal averaged contrast values, calculated on a linear scale for each dark region, are $6.8 \times 10^{-8}$, $8.0 \times 10^{-8}$, $7.0 \times 10^{-8}$ and $4.7 \times 10^{-8}$for DR 1, DR 2, DR 3 and DR 4, respectively, where DR 1 $\sim$ DR 4 are the dark regions corresponding to the quadrants around the core shown in Fig. 1. A value of $6.7 \times 10^{-8}$was obtained by averaging in the same way, but for all of the DRs. A profile of the radial average, calculated on a linear scale for all of four DRs, is also shown in Fig. 6. It is possible to define the distance beyond which the contrast falls below 10-6, IWA6, for each dark region. Thus, IWA6 was $3.4~\lambda/D$ for DR 1, DR 3, DR 4, and was $3.6~\lambda/D$ for DR 2. IWA6 for the profile of the radial average was $3.4~\lambda/D$. These results are close to the theoretical value of 3.2 $\lambda/D$.

Figure 4b exhibits an irregular pattern of flux distribution in the dark region, which is not predicted by the theoretical PSF for the mask. It was confirmed that the shape of this irregular pattern was approximately repeatable in a fixed setup configuration. The same measurement was performed, with and without clean air flow or air suspension for the optical table, to test the influence of air turbulence or vibration of the devices, and confirmed that no significant change occurred. In the experiment in Paper I, tests were performed, rotating and shifting the mask parallel to the mask surface, to confirm whether the intensity distribution of the irregular pattern changed. Such a test provided useful information to ascertain the limiting factor in the coronagraph performance. However, it was difficult to move the checkerboard mask in this work in the present optical setup because the mechanics is complicated by defects of the mask pattern and the cover mask, requiring a fine tuning stage for the cover mask. Therefore, to achieve such an examination by moving the mask in the beam, all the optics were shifted, except the devices fixed to pupil mask (i.e., checkerboard mask, cover mask, fine tuning stage and baffle with it). The extent of the shift was 3 mm, and the direction of the shift was parallel to the optical table and perpendicular to the optical axis of the laser. It was confirmed that the shape and intensity distribution of the irregular pattern was changed. Hence it is concluded that errors in the pupil shaping were not a major limiting factor in the coronagraphic performance. We consider that speckles are the primary limiting factor, caused by wavefront errors produced by imperfection of the surface of optics, multiple reflections by the transmissive optical device, scattering by microscopic defects on the surface of the optics, stray light, and so on. In the dark region from Mask 3, we could not find the lattice pattern that had been observed in the outer part of the dark region from the Mask 2 coronagraph, shown in Paper I. This result is reasonable because the lattice pattern is expected to appear when the measurement accuracy reaches the design limit, and the designed contrast was 10-10 for Mask 3 whilst it was 10-7 for Mask 2.

In order to estimate the optical bench stability, we performed a series of 10, 1 h exposure time acquisitions. Consecutive exposures were compared by simply subtracting them, after some basic image processing, as described above. The speckle fluctuation in the entire dark region for all of the 10 images is estimated to be $\sim$ $1.5\times10^{-8}$ rms, compared to the peak intensity of the PSF core. Within a $5~\times~5~(\lambda/D)^2$ area, closer to the optical axis, these average fluctuations were $\sim$ $3.1\times10^{-8}$. We relate these fluctuations to changes of temperature, which could have affected the mechanical stability of the optical bench during the 10 h long observation session. On the best two images continually obtained, the intensity fluctuations were of the order of $6.8 \times 10^{-9}$. If rolling of the telescope were possible during observation, these quasi-static speckles caused by imperfection of the whole optics including the telescope, would not change, whilst a planet would rotate in the field. The ``roll subtraction'' would therefore exhibit a typical plus-minus signature if an object were located in the field. We computed that a rotation angle of about 8 degrees would be required to detect a planet at 6$\lambda/D$ distant from the central star so that its intensity would not be reduced by more than 50\% by the roll subtraction. This also puts additional constraints on the minimal distance at which such a detection can be performed if the roll subtraction is used. With the numbers given above, the direct detection of a planet with a contrast of 10-7 would be possible with a signal-noise ratio above 10 by the roll subtraction of two images.

The Space Infrared telescope for Cosmology and Astrophysics (SPICA) is the next generation mission for infrared astronomy, led by Japan (Nakagawa et al. 2004). A mid-infrared coronagraphic instrument to provide 10-6contrast is currently being considered for the SPICA mission for the direct observation of Jovian extra-solar planets (Enya et al. 2007b; Abe et al. 2007; Enya et al. 2006; Tamura 2000). Adaptive wavefront correction is one of the candidate methods to correct wavefronts from the telescope and to achieve 10-6 contrast. However, development of a cryogenic deformable mirror for the telescope cooled to 4.5 K is challenging. If the rigid coronagraph with PSF subtraction can provide 10-6 contrast, the system will be quite simple, feasible and reliable. Thus, extensive estimation of the quality and stability of whole system and trade-offs of specifications are needed to determine which method is suitable for SPICA, although the quality and stability of not only the coronagraphic part, but the whole system of the spacecraft, is important for the rigid coronagraph mission. SEE-COAST is a space mission concept (Schneider et al. 2006), submitted to the European Space Agency (ESA Cosmic Vision 2015-2020). The target contrast of SEE-COAST is 10-8 and a type of rigid coronagraph is being considered for use in this mission. Therefore, the result demonstrated in this work supports the design concept of SEE-COAST. A rigid coronagraph is also suitable for small satellites, because the limitations of weight, volume, and power are more serious in small satellites. In principle, combination of a rigid coronagraph and PSF subtraction works regardless of the coronagraphic method. However, the binary masks in general offer extremely good stability especially for the tip-tilt jitter sensitivity, and a coronagraph with a binary shaped pupil mask would be suitable coronagraphic method for a rigid coronagraph.

\section{Conclusion}
We present the results of a laboratory experiment intended to produce a rigid coronagraph in air, without adaptive wavefront correction. A new checkerboard pupil mask, Mask 3 was designed and manufactured, and the optics for the experiment are improved from Paper I. Precise evaluation of coronagraphic performance, including PSF subtraction, has been carried out. As the result, a contrast of $6.7 \times 10^{-8}$ was achieved for the raw coronagraphic image by areal averaging of all of the observed dark regions. Following the PSF subtraction, the contrast reached $6.8 \times 10^{-9}$. It is concluded that speckles are the major limiting factor over the entire dark region of both the raw image and the PSF subtracted image. A rigid coronagraph with PSF subtraction is a useful method to achieve high contrast observations, whilst adaptive wavefront correction is another strong tool. A trade-off study is important for each coronagraph mission to determine which method should be adopted: either a coronagraph with adaptive wavefront correction, or a rigid coronagraph.

\begin{acknowledgements}
We are grateful to N. Nakagiri and other colleagues at AIST for their support in the mask fabrication. We would like to thank N. Okada and Advanced Technology Center of National Astronomical Observatory Japan. We also thank all involved in the to SPICA mission. This work was supported in part by a grant from the Japan Science and Technology Agency. LA is supported by Grants-in-Aid (No. 160871018002) from the Ministry of Education, Culture, Sports, Science, and Technology of Japan.
\end{acknowledgements}


\clearpage
 \begin{figure}
   \centering
\vspace*{70mm}
   \includegraphics[height=6cm]{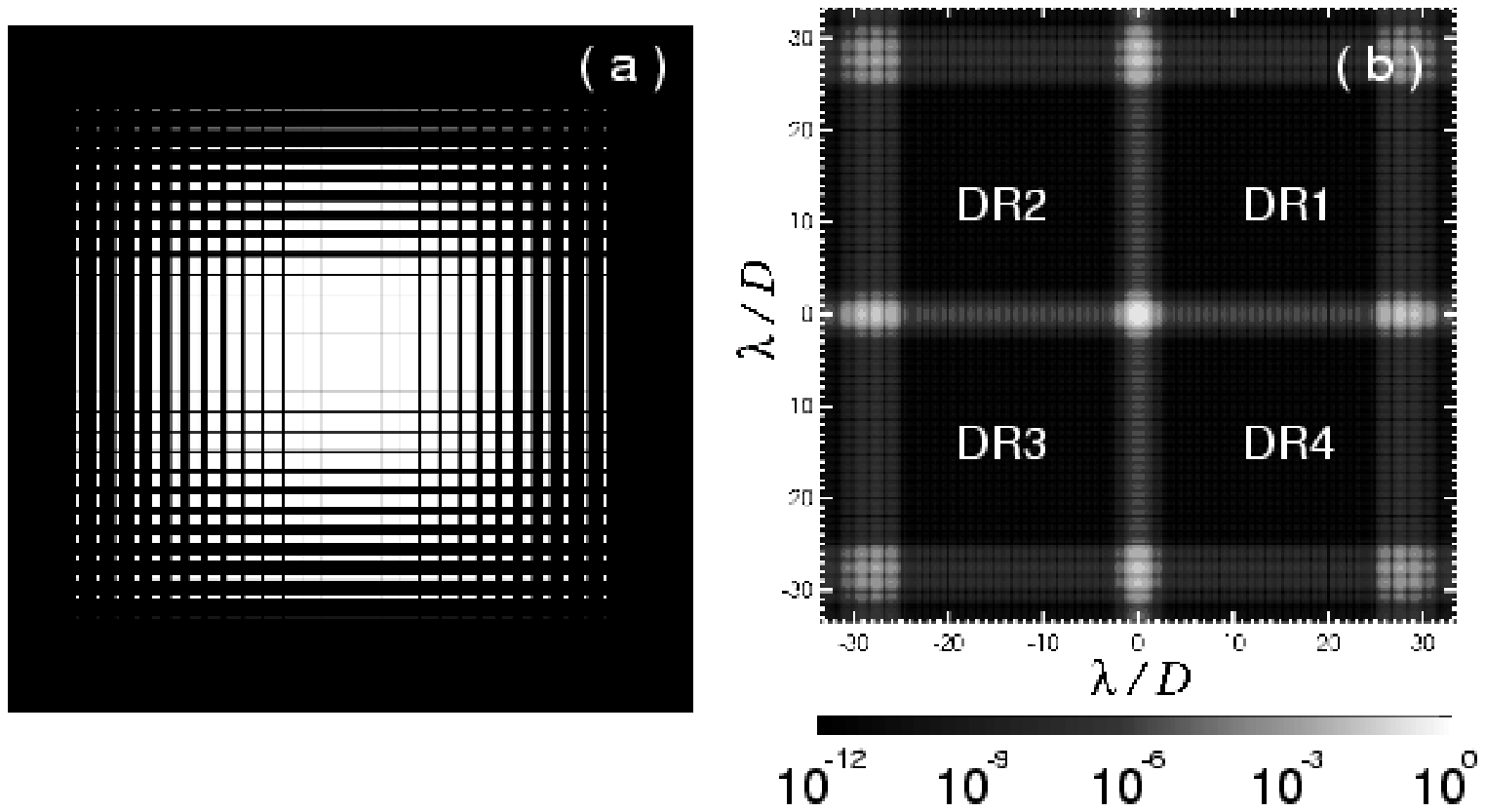}
   \caption{anel  a) Shape of the mask. The transmissivities of the black and white regions are 0 and 1, respectively. Diameter of circumscribed circle to the transmissive region is 2 mm. Panel  b) Simulated coronagraphic PSF, using the mask design shown in panel  a). The profile of the designed PSF is shown in Fig. 5. Four dark regions in panel  b) are named DR 1, DR 2, DR 3, and DR 4, respectively. }
   \label{fig1}
   \end{figure}

\clearpage
 \begin{figure}
   \centering
\vspace*{70mm}
   \includegraphics[height=10cm]{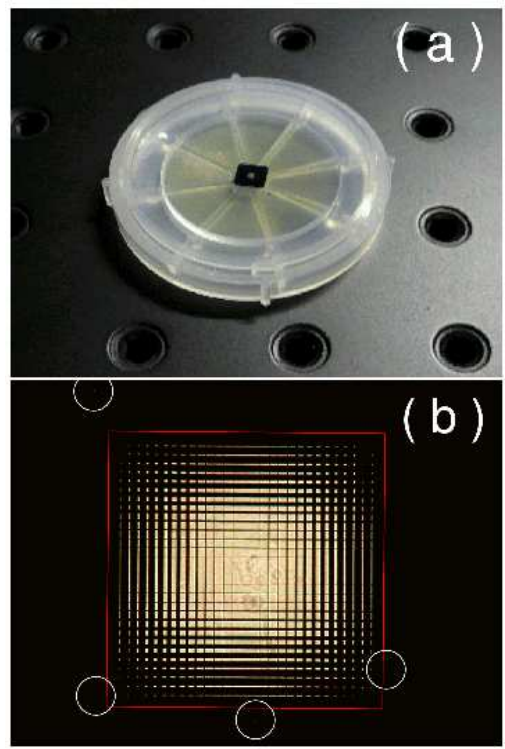}
   \caption{Panel  a) Fabricated mask on a BK 7 substrate. The diameter and thickness of the substrates are 30.0 mm and 2.0 mm, respectively. Panel  b) Transmission microscope photograph of the fabricated mask. White circles indicate positions of small defects (holes) in the aluminium film. The red square corresponds to size of the square hole of the cover mask. }
   \label{fig1}
   \end{figure}

\clearpage
 \begin{figure}
   \centering
\vspace*{70mm}
   \includegraphics[height=10cm]{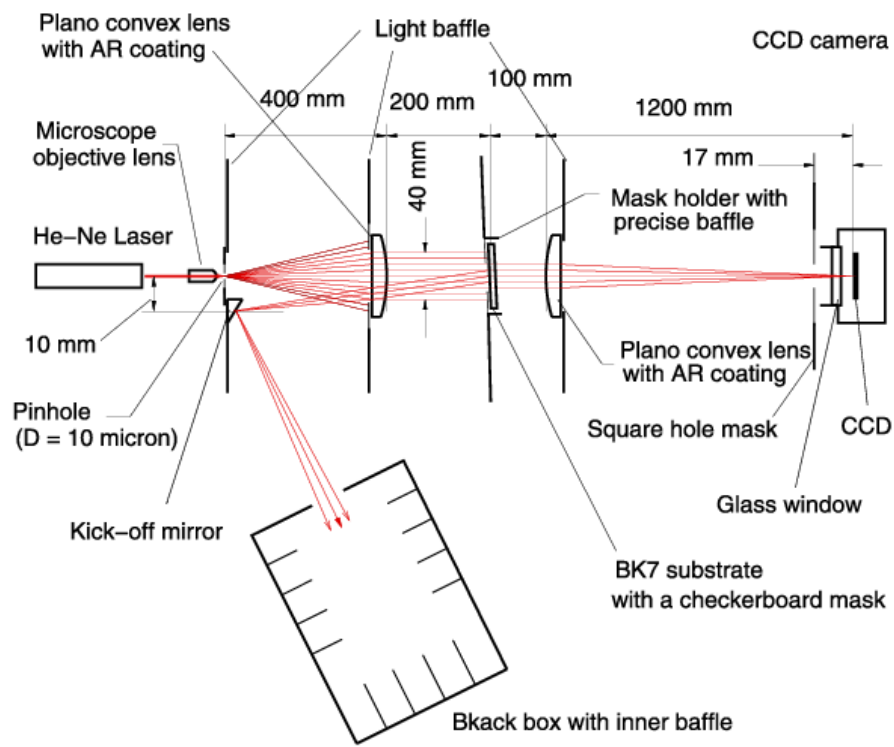}
   \caption{Configuration of the experimental optics. The entire optical device was set on an optical bench with air suspension, in a dark room with a clean-air flow system. }
   \label{fig1}
   \end{figure}

\clearpage
 \begin{figure}
   \centering
\vspace*{70mm}
   \includegraphics[height=12cm]{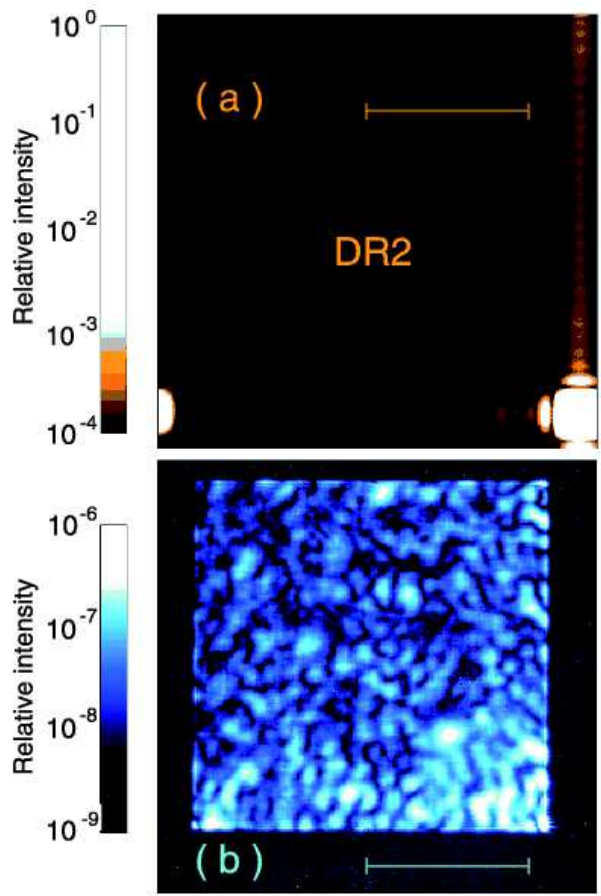}
   \caption{bserved coronagraphic images. Panel  a) shows images including the core of the PSF. The perpendicular tail over the bright peak is along the readout direction. Panel  b) is the image of a dark region obtained with a blank square mask. Scale bar corresponds to $10 \lambda / D$. }
   \label{fig1}
   \end{figure}

\clearpage
 \begin{figure}
 \vspace*{70mm}
  \centering
   \includegraphics[height=8cm]{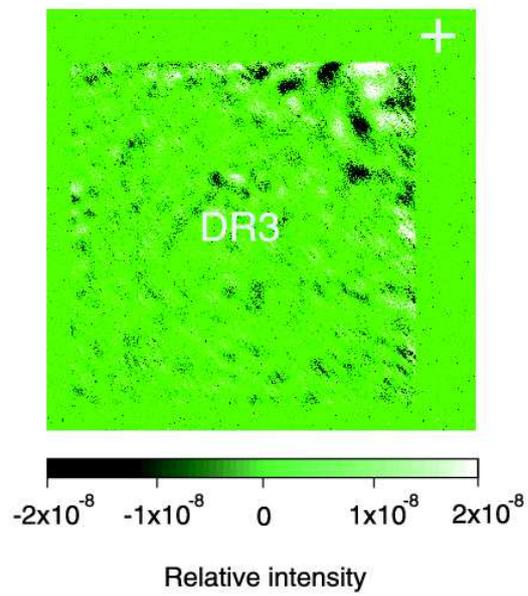}
   \caption{esult of PSF subtraction using two images of DR 3 with a 3600 s exposure. A cross shows the position of the PSF center. }
   \label{fig1}
   \end{figure}

\clearpage
 \begin{figure}
\vspace*{70mm}
   \centering
   \includegraphics[height=10cm]{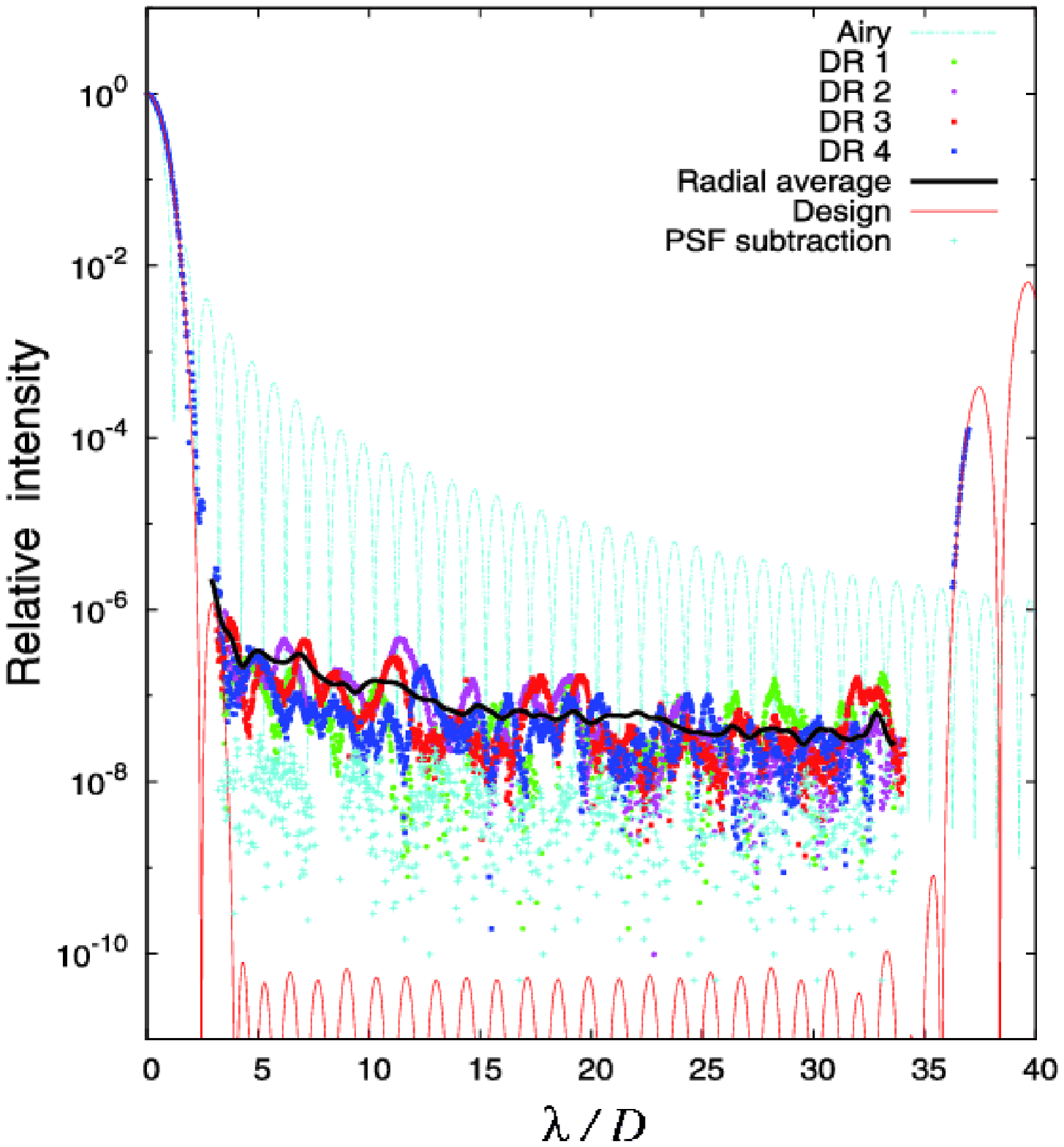}
   \caption{Diagonal profile of the coronagraphic image. Observed and designed profiles are shown with an Airy profile. Peak intensity of each profile is normalized. The radial average of observed 4 dark regions, calculated on a linear scale, is also presented. }
   \label{fig1}
   \end{figure}


\begin{thebibliography}{}

\bibitem[2007]{Abe}
Abe, L., Enya, K., Tanaka, S., et al. 2007, Optical techniques for direct imaging of exoplanets, Comptes Rendus Physique, 8, 374


\bibitem[2006]{Belikov}
Belikov, R., Give'on, A., Trauger, J. T., et al. 2006, Proc. SPIE, 6265, 626518 

\bibitem[2004]{Burrows}
Burrows, A., Sudarsky, D., \& Hubeny, I. 2004, ApJ, 609, 407 

\bibitem[2006]{Enya}
Enya, K., Tanaka, S., Nakagawa, T., et al. 2006, Proc. SPIE, 6265, 626536

\bibitem[2007a]{Enya}
Enya, K., Tanaka, S., Abe, L., \& Nakagawa, T. 2007a, A\&A, 461, 783

\bibitem[2007b]{Enya}
Enya, K., Abe, L., Tanaka, S., et al. 2007b, Proc. SPIE, 6693, in press 

\bibitem[2004]{Green}
Green, J. J., Shaklan, S. B., Vanderbei, R. J., \& Kasdin, N. J. 2004, Proc. SPIE, 5487, 1358 

\bibitem[1964]{Jacquinot}
Jacquinot, P., \& Roizen-Dossier, B. 1964, Prog. Opt., 3, 29 

\bibitem[2003]{Kasdin}
Kasdin, N. J., Vanderbei, R. J., Spergel, D. N., \& Littman, M. G. 2003, ApJ, 582, 1147 

\bibitem[1936]{Lyot}
Lyot, B. 1936, MNRAS, 99, 580

\bibitem[2004]{Nakagawa}
Nakagawa, T., \& SPICA working group 2004, Adv. Space Res., 34, 645 

\bibitem[2006]{Schneider}
Schneider, J., Riaud, P., Tinetti, G., et al. 2006, The See-Coast Team Proc. of the Annual meeting of the French Society of Astronomy and Astrophysics, ed. D. Barret, F. Casoli, G. Lagache, A. Lecavelier, \& L. Pagani, 429 

\bibitem[2001]{Spergel}
Spergel, D. N. 2001 [arXiv:asro-ph/0101141] 

\bibitem[2000]{Tamura}
Tamura, M. 2000, ISAS Rep., 14 

\bibitem[2006]{Tanaka}
Tanaka, S., Enya, K., Abe, L., Nakagawa, T., \& Kataza, H. 2006, PASJ, 58, 627 

\bibitem[2007]{Trauger}
Trauger, J. T., \& Traub, W. A. 2007, Nature, 446, 771  

\bibitem[1999]{Vanderbei}
Vanderbei, R. J. 1999, Optimization methods \& software, 21, 485 

\bibitem[2004]{Vanderbei}
Vanderbei, R. J., Kasdin, N. J., \& Spergel, D. N. 2004, ApJ, 615, 555 




\end{thebibliography}
\end{document}